\documentclass[showpacs,twocolumn,prx,superscriptaddress,notitlepage]{revtex4-1}

\usepackage{qcircuit}
\usepackage[dvips]{graphicx}
\usepackage{amsmath,amssymb,amsthm,mathrsfs,amsfonts,dsfont}
\usepackage{subfigure, epsfig}
\usepackage{braket}
\usepackage{bm}
\usepackage{enumerate}
\usepackage{soul, color}
\usepackage{fancybox, graphicx}
\usepackage[skins]{tcolorbox}
\usepackage{chemformula}
\usepackage{physics}

\newtcolorbox[auto counter]{tbox}[2][]{%
    enhanced, float=hbt, drop fuzzy shadow southeast,
    colback=white!5!white, colframe=white!50!black,
    width= .97\columnwidth,sharp corners,boxrule=0.8pt,
    title={Algorithm \thetcbcounter: #2}, #1
}

\newcommand{\p}{\mathbf{ p}}

\newcommand{\q}{\mathbf{q}}
\newcommand{\darkd}{\mathbf{d}}
\newcommand{\Q}{\mathbf{q}}

\newcommand{\creat}{{a}^\dag}
\newcommand{\annih}{{a}}
\newcommand{\Hamilt}{{H}}
\newcommand{\Qsingle}{q}
\newcommand{\U}{\hat{U}}
\newcommand{\red}[1]{\textcolor{black}{#1}}

\newtcolorbox{codebox}{enhanced,width=.95\columnwidth, halign = flush left, drop fuzzy shadow southeast,boxrule=0.4pt,sharp corners,colframe=black,colback=white}

\begin{document}

\title{\red{Digital quantum simulation of molecular vibrations}}

\begin{abstract}
Molecular vibrations underpin important phenomena such as spectral properties, energy transfer, and molecular bonding. However, obtaining a detailed understanding of the vibrational structure of even small molecules is computationally expensive. While several algorithms exist for efficiently solving the electronic structure problem on a quantum computer, there has been comparatively little attention devoted to solving the vibrational structure problem with quantum hardware. In this work, we discuss the use of quantum algorithms for investigating both the static and dynamic vibrational properties of molecules. We introduce a physically motivated unitary vibrational coupled cluster ansatz, which also makes our method accessible to noisy, near-term quantum hardware. We numerically test our proposals for the water and sulfur dioxide molecules.
\end{abstract}

\date{\today}
\author{Sam McArdle}
\affiliation{Department of Materials, University of Oxford, Parks Road, Oxford OX1 3PH, United Kingdom}

\author{Alexander Mayorov}
\affiliation{Department of Materials, University of Oxford, Parks Road, Oxford OX1 3PH, United Kingdom}
\affiliation{Department of Chemistry, University of Cambridge, Lensfield Road, Cambridge CB2 1EW, United Kingdom}

\author{Xiao Shan}
\affiliation{Physical and Theoretical Chemical Laboratory, University of Oxford, South Parks Road, Oxford OX1 3QZ, United Kingdom}

\author{Simon Benjamin}
\affiliation{Department of Materials, University of Oxford, Parks Road, Oxford OX1 3PH, United Kingdom}

\author{Xiao Yuan}
\email{xiao.yuan.ph@gmail.com}
\affiliation{Department of Materials, University of Oxford, Parks Road, Oxford OX1 3PH, United Kingdom}

\maketitle

\section{Introduction}

Simulating many-body physical systems enables us to study chemicals and materials without fabricating them, saving both time and resources. The most accurate simulations require a full quantum mechanical treatment --- which is exponentially costly for classical computers. While many approximations have been developed to solve this problem, they are often not sufficiently accurate~\cite{helgaker2014molecular}. One possible route to more accurate simulations is to use quantum computers. Quantum computation can enable us to solve certain problems asymptotically more quickly than with a `classical' computer~\cite{feynman1982simulating, shor1994algorithms, grover1996fast}. While the quantum computers that we currently possess are small and error prone, it is hoped that we will one day be able to construct a universal, fault-tolerant quantum computer --- widely expected to be capable of outperforming its classical counterparts on certain tasks. One example of such tasks is simulating quantum systems on quantum computers~\cite{lloyd1996universal, Abrams97, Abrams99}. In particular, simulating chemical systems, such as molecules~\cite{aspuru2005simulated}, has received significant attention. This may stem from the commercial benefits of being able to investigate and design such systems \textit{in silico}~\cite{Revolution}. The development of quantum computational chemistry has arguably echoed its classical counterpart. In both fields, the majority of investigations have focused on the electronic structure of molecules~\cite{christiansen2007vcc}. This has resulted in a wealth of well established methods for solving problems of electrons. However, methods concerned with the nuclear degrees of freedom are comparatively less well established. Understanding vibrations is critical for obtaining the most accurate models of real physical systems~\cite{christiansen2007vcc}. Unfortunately, the most detailed classical simulations of vibrations are limited to small molecules, consisting of a few atoms~\cite{bowman2008variational}. While approximations can be used to treat larger systems, these tend to be less accurate than experiments~\cite{christiansen2004secondquantised}. Although recently proposed analog quantum algorithms~\cite{huh2015boson,huh2017vibronic,clements2017experimental,shen2018vibronic, sparrow2018simulating,chin2018quantum,hu2018simulation} are capable of simulating molecular vibrations using resources which scale polynomially with the size of the molecule, the long term scalability of these approaches has yet to be established. 

In this work, we discuss a general method for efficiently simulating molecular vibrations on a universal quantum computer. Our method targets the eigenfunctions of a vibrational Hamiltonian with potential terms beyond quadratic order (`anharmonic potentials'). These wavefunctions can then be used to efficiently calculate properties of interest, such as absorption spectra at finite temperatures. We can also use our method to perform simulations of vibrational dynamics, enabling the investigation of properties such as vibrational relaxation.

\section{Vibrations}
A consequence of quantum mechanics is that molecules are never at rest, possessing at least the vibrational zero-point energy correction to the ground-state energy~\cite{ZPEdynamics, ZPESiH, Gross1997}. As a result, vibrations affect all chemical calculations, to a greater or lesser extent. They are important in both time dependent and independent contexts. From a dynamics perspective, vibrational structure affects high frequency time-resolved laser experiments~\cite{seideman2000spectroscopy}, reaction dynamics~\cite{VibrExcitation, ProteinVibr, Proctor2008}, and transport~\cite{ElectronTransferRates1, ElectronTransferRates2}. In a static context, vibrations underpin spectral calculations, such as: infrared and Raman spectroscopy~\cite{FCFApplications} and fluorescence~\cite{GFPModelling}. These calculations determine the performance of solar cells~\cite{yue2016solar,debbichi2012solar} and industrial dyes~\cite{dhananasekaran2016dye,biswas1997dye}, as well as the susceptibility of molecules to photodamage~\cite{choi2005photodamage,huh2015boson}.

Despite their importance for accurate results, studying vibrations has proven difficult. There are several possible routes to obtaining an accurate description of vibrational behaviour. Real-space, grid based methods, which treat the electronic and nuclear degrees of freedom on an equal footing, are limited to systems of a few particles. While algorithms to efficiently solve this problem on a universal quantum computer exist~\cite{kassal2008polynomial,KivReal}, it will take many years to develop a quantum computer with the required number of qubits~\cite{jones2012faster}. Alternatively, one may separate the electronic and nuclear degrees of freedom. We can solve for the electronic energy levels of the system as a function of the nuclear positions, which enables us to map out potential energy surfaces for the system. A number of approximate classical methods have been developed to solve this problem~\cite{helgaker2014molecular,szabo2012modern}, as well as several quantum algorithms~\cite{aspuru2005simulated,babbush2016exponential,babbush2017exponential,peruzzo2014variational}. These electronic potential surfaces can then be viewed as the nuclear potential, determining the vibrational energy levels. This is known as the vibrational structure problem. The accuracy of the nuclear potential is determined by the accuracy of the electronic structure calculation, as well as the number of points obtained for the potential energy surface. Once this potential has been obtained, a number of classical methods can be used for solving both the time dependent and independent Schr\"odinger equations.

The most simple methods \red{uses} the `harmonic approximation'. This treats the nuclear potential in the vicinity of the equilibrium geometry as a harmonic oscillator potential, resulting in energy eigenstates which are harmonic oscillator eigenfunctions. 

Alternatively, one may consider higher order expansions of the nuclear potential, resulting in more accurate calculations~\cite{christiansen2007vibrationalstructure}. 
\red{One common route towards obtaining the nuclear potential is to first carry out many electronic structure calculations on the system, in the vicinity of the minimum energy configuration. Each of these electronic structure calculations is approximate, and so the cost of each one scales polynomially with the system size. However, if one proceeds to obtain the nuclear potential using this simple grid based method, then a number of grid points scaling exponentially with the number of modes is required~\cite{christiansen2012selected}. In practice one can often instead construct an approximate nuclear potential by considering a reduced number of mode couplings, or using interpolation, or using adaptive methods. A review of these, and other state-of-the-art methods can be found in Ref.~\cite{christiansen2012selected}.
The requirement to first perform multiple electronic structure calculations to obtain the anharmonic nuclear potential makes calculating vibrational energy levels expensive~\cite{huhthesis}, even if only mean-field vibrational calculations are then performed. If the correlation between different vibrational modes is included in the calculation, then the simulation becomes even more expensive. While most of the existing classical vibrational simulation methods scale polynomially with the number of modes in the system (e.g. vibrational self-consistent field methods~\cite{christiansen2004secondquantised}, or vibrational coupled cluster theory~\cite{christiansen2004vcc}), and are sufficiently accurate for some systems, they only provide approximations to the true full configuration interaction vibrational wavefunction, which can be exponentially costly to obtain. A similar hierarchy of accuracy also exists for dynamics simulations.}

The computational difficulties described above make accurate vibrational calculations on large systems very challenging for classical computers. To overcome these challenges, quantum solutions have been suggested for the vibrational structure problem \cite{joshi2014francknmr,huh2015boson,huh2017vibronic,clements2017experimental,shen2018vibronic, sparrow2018simulating,chin2018quantum,hu2018simulation}. To date, the majority of suggestions have focused upon analog quantum simulation of vibrations. In analog simulations, the simulator emulates a specific system of interest, but cannot in general be programmed to perform simulations of other, different systems. Huh~\textit{et al.} proposed using boson sampling circuits to determine the absorption spectra of molecules~\cite{huh2015boson}. These boson sampling circuits consist of photons passing through an optical network. This initial proposal relied on the harmonic oscillator approximation at zero temperature, but does take into account bosonic mode mixing due to nuclear structural changes that result from electronic excitation. This method has since been experimentally demonstrated~\cite{shen2018vibronic,clements2017experimental}, and extended to finite temperature spectra~\cite{huh2017vibronic,hu2018simulation}. The main limitation of these simulations is the use of the harmonic oscillator approximation for the vibrational wavefunction. It is in general difficult to engineer ground states of anharmonic Hamiltonians using an optical network, as non-linear operations, such as squeezing, are required. Optical networks have also been used for simulating vibrational dynamics~\cite{sparrow2018simulating}. These simulations investigated vibrational transport, adaptive feedback control, and anharmonic effects.

The aforementioned schemes make use of the analogy between the vibrational energy levels in molecules in the Harmonic oscillator approximation, and the bosonic energy levels accessible to photons and ions. One advantage of this is that the bosonic modes are in principle able to store an arbitrary number of excitations. As these analog simulators are relatively simple to construct (when compared to a universal, fault-tolerant quantum computer), they will likely prove useful for small calculations in the near-term. However, it is not yet known how to suppress errors to an arbitrarily low rate in analog simulators. As a result, if we are to simulate the vibrational behaviour of larger quantum systems, we will likely require error corrected universal quantum computers. This motivates our work on methods for vibrational simulation on universal quantum computers.

The rest of this paper is organised as follows. 
In Sec.~\ref{Sec:encoding}, we introduce the vibrational stucture problem for molecules and show how this problem can be mapped onto a quantum computer. In Sec.~\ref{Sec:vibrations}, we show how to solve both static and dynamic problems of molecular vibrations. Finally, in Sec.~\ref{Sec:simulation}, we present the results of numerical simulations of the \ch{H2O} and \ch{SO2} molecules.

\section{Encoding}\label{Sec:encoding}
\paragraph{Vibrational Hamiltonian.}

Under the Born-Oppenheimer approximation,
\red{nuclear variables are treated as parameters in the electronic structure problem and are restored as quantum nuclear variables at the level of the full problem.}
In the following, we neglect the rotational degrees of freedom \red{from negligible rotational-vibrational couplings for rigid molecules (rigid rotator approximation)}. 
After diagonalising the electronic Hamiltonian \red{and neglecting nonadiabatic couplings}, the molecular Hamiltonian becomes
\begin{equation}
	\Hamilt_{mol} = \sum_s\ketbra{\psi_s}_e\otimes H_s,
\end{equation}
where $\ket{\psi_s}_e$ are the electronic energy eigenstates. The effective nuclear Hamiltonian \red{$H_s$}  is
\begin{equation}
	\Hamilt_s = \frac{\p^2}{2} + V_s(\Q),
\end{equation}
where $\Q = (q_1, q_2\dots)$ are nuclear coordinates, $\p=-i\partial/\partial \Q$ are nuclear momenta \red{with $\hbar=1$}, and $V_s(\Q)$ is the effective nuclear potential. This potential is determined by the corresponding electronic potential energy surface of $\ket{\psi_s}_e$. As described in Appendix~\ref{ProblemMapping}, we work in mass-weighted normal coordinates and decouple the rotational and vibrational modes. The potential $V_s(\Q)$ can be approximated as
\begin{equation}
V_s(\Q) \approx \frac{1}{2}\sum_i \omega_i^2 \Qsingle_i^2,
\end{equation}
where $\omega_i$ is the harmonic frequency of the $i$\textsuperscript{th} vibrational normal mode.
Thus, the nuclear Hamiltonian $\Hamilt_s$ can be approximated by a sum of independent harmonic oscillators,
\begin{equation}
	\Hamilt_s \approx \sum_i \omega_i\creat_i \annih_i,
\end{equation}
with $\creat_i$ and $\annih_i$ being the creation and annihilation operator of the $i$\textsuperscript{th} harmonic oscillator. 
This is the commonly used `harmonic approximation'. Even for accurate potentials and rigid molecules, the harmonic approximation is less accurate than modern spectroscopic techniques~\cite{christiansen2004secondquantised}. This approximation becomes inadequate for large and `floppy' molecules~\cite{christiansen2004secondquantised}. Improved results can be obtained by including anharmonic effects which requires information of higher order potential terms in the Hamiltonian~\cite{christiansen2012selected}. For example, we can expand the potential as
\begin{equation}
	V_s(\Q) = \sum_{j=2}^\infty \sum_{i_j} k_{i_1, i_2, \dots, i_j} q_{i_1}q_{i_2}\dots q_{i_j},
\end{equation}
where $k_{i_1, i_2, \dots, i_j} $ are the coefficients of the expansion $q_{i_1}q_{i_2}\dots q_{i_j}$, and the harmonic frequencies are $\omega_i = \sqrt{2k_{i,i}}$. In general, the eigenstates of these Hamiltonians are entangled states, when working in a basis of harmonic oscillator eigenstates. Consequently, solving the higher order vibrational Hamiltonian is a hard problem for classical computers.

In contrast, we show below that it is possible to efficiently encode the $k$\textsuperscript{th} order nuclear Hamiltonian into a Hamiltonian acting on qubits. We can then use quantum algorithms to efficiently calculate the static and dynamic properties of the nuclear Hamiltonian.\\

\paragraph{Mapping to qubits}
We first discuss mapping the molecular Hamiltonian into qubits. We work in the basis of harmonic oscillator eigenstates, as these can be easily mapped to qubits. The direct mapping presented below was originally suggested in the context of simulating general bosonic systems by Somma~\textit{et al.}~\cite{somma2003simulating2}. It has been used recently in the context of quantum simulation of nuclear physics to investigate the binding energy of a deuteron nucleus~\cite{dumitrescu2018cloud}. The compact mapping discussed below was proposed by Veis~\textit{et al.}, in the context of using quantum computers to simulate `nuclear orbital plus molecular orbital (NOMO)' theory, which uses Gaussian orbitals for the nuclei, and treats them on an equal footing to the electrons~\cite{veis2016beyond}. This differs from our work, which separates the nuclear and electronic degrees of freedom, and predominantly considers a harmonic oscillator basis for the vibrational modes. While this tailors our method for vibrational problems, it means we are limited to solving problems for which the Born-Oppenheimer approximation is valid, unlike Ref.~\cite{veis2016beyond}.\\

Focusing first on one harmonic oscillator, $\hat{h} = \omega \creat \annih$, we consider the truncated  eigenstates with the lowest $d$ energies, $\ket{s}$ with $s=0,1,\dots,d-1$. A \textit{direct mapping} of the space $\{\ket{s}\}$ is to encode it with $d$ qubits as
\begin{equation}
	\ket{s} = \otimes_{j=0}^{s-1}\ket{0}_{j}\ket{1}_{s}\otimes_{j=s+1}^{d-1}\ket{0}_{j},
\end{equation}
with creation operator 
\begin{equation}
	\creat = \sum_{s=0}^{d-2}\sqrt{s+1}\ketbra{0}{1}_{s}\otimes\ketbra{1}{0}_{s+1}.
\end{equation}
The annihilation operator can be obtained by taking the Hermitian conjugate of $\creat$.
As an alternative to the direct mapping, we can use a \textit{compact mapping}, using $K = \lfloor\log d\rfloor$ qubits,
\begin{equation}
	\ket{s} = \ket{b_{K-1}}\ket{b_{K-2}}\dots \ket{b_0},
\end{equation}
with binary representation $s = b_{K-1}2^{K-1}+b_{K-2}2^{K-2}+\dots b_02^{0}$. The representation of the creation operator is
\begin{equation}
	\creat = \sum_{s=0}^{d-2}\sqrt{s+1}\ketbra{s+1}{{s}}.
\end{equation}

These binary projectors can then be mapped to Pauli operators;
\begin{equation}
\begin{aligned}
	\ket{0}\bra{0} = \frac{1}{2}(I + Z),\,&
	\ket{1}\bra{1} = \frac{1}{2}(I - Z),\\
	\ket{1}\bra{0} = \frac{1}{2}(X - iY), \,&
	\ket{0}\bra{1} = \frac{1}{2}(X + iY).
\end{aligned}
\end{equation}

\begin{figure}[t]\centering
  \includegraphics[width=1\linewidth]{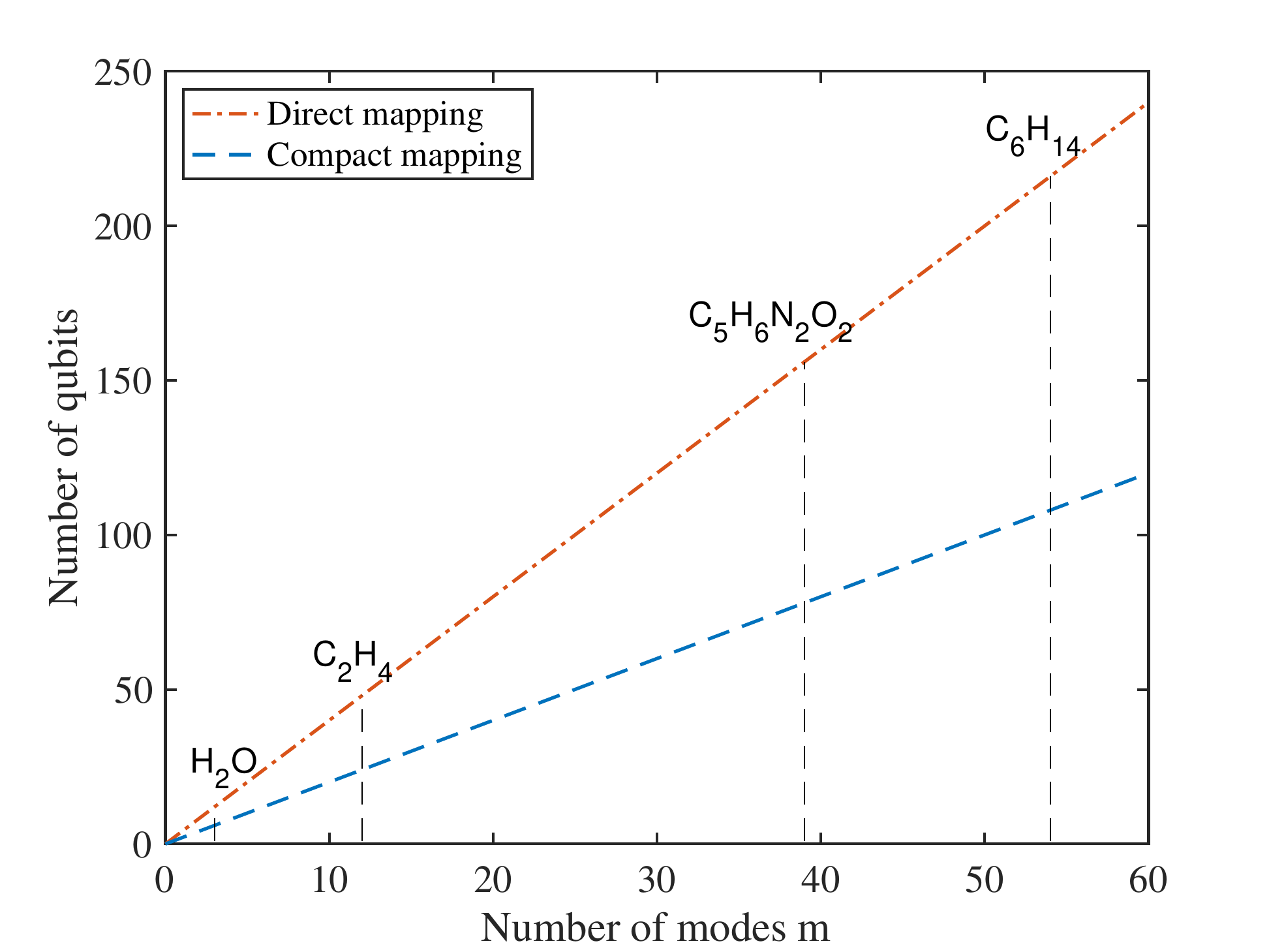}
  \caption{Number of qubits required for the direct and compact mappings with $d = 4$ energy levels for each mode.}\label{Fig:qubits}
\end{figure}

When decomposing $\creat$ and $\annih$ into local Pauli matrices, there are $O(d)$ and $O(d^2)$ terms for the direct and compact mapping, respectively. In Fig.~\ref{Fig:qubits}, we show the number of qubits required to describe the vibrational Hamiltonians of several molecules, for both mappings.

As $\p$ and $\q$ can both be represented by a linear combination of creation and annihilation operators, we can thus map the nuclear vibrational Hamiltonian to a qubit Hamiltonian. If the molecule has $n$ atoms, it has $M = 3n-6$ vibrational modes for a nonlinear molecule, and $M = 3n-5$ for a linear molecule. The vibrational wavefunction can then be represented with $Md$ (direct mapping) or $M\log (d)$ (compact mapping) qubits. This can be contrasted with the exponentially scaling classical memory required to store the wavefunction. If the potential is expanded to $k$\textsuperscript{th} order (with $M\gg k$), the Hamiltonian contains $O(M^k d^k)$ (direct) or $O(M^k d^{2k})$ (compact) terms. 
These terms are strings of local Pauli matrices. In this work, \red{we take $d$ to be a small constant. This approximation constrains us to the low energy subspace of the Hamiltonian, which should be valid for calculations of ground and low-lying excited states. The applicability of this approximation to the simulation of dynamics is discussed in Sec.~\ref{Sec:Discussion}.}
We set $k = 4$ to investigate the Hamiltonian to quartic order. The resulting Hamiltonian has $O(M^4)$ terms.

\section{Simulating molecular vibrations}\label{Sec:vibrations}
Once the vibrational modes have been mapped to qubits, we can use quantum algorithms to obtain the static and dynamic properties of the system. We can write the qubit Hamiltonian as $\Hamilt_s = \sum_i \lambda_i {h}_i$, where $h_i$ are coefficients determining the strength of each term in the Hamiltonian.

\subsection{Vibrational energy levels}
An important, but classically difficult problem, is to obtain accurate energy levels for the vibrational Hamiltonian. The spectrum of the vibrational Hamiltonian provides corrections to the electronic eigenstates used to predict reaction rates~\cite{Gross1997}. Moreover, we will show how these energy levels can be used to calculate the absorption spectrum of molecules in Sec.~\ref{FCFSec}. Of particular interest are the lowest lying energy levels at low temperature. Using a universal quantum computer, we can first prepare an initial state that has a large overlap with the ground state of the vibrational Hamiltonian. We can then use the phase estimation algorithm~\cite{kitaev1995phase,Abrams99} to probabilistically obtain the ground state and ground state energy. A possible initial ground state is the lowest energy product state of the harmonic oscillator basis states, $\ket{\psi_0} = \otimes_m\ket{s_m}_m$. However, we note that the overlap between this state and the true ground state may decrease exponentially with the size of the molecule. This so-called `orthogonality catastrophe' has been discussed previously in the context of electronic structure calculations on a quantum computer~\cite{mcclean2014locality,tubman2018orthogonality}. As a result, for large systems it may be more efficient to use an initial state obtained from a classical vibrational self-consistent field (VSCF) calculation. VSCF is the vibrational analogue of the Hartree-Fock method in electronic structure theory. VSCF optimises the basis functions to minimise the energy of the Hamiltonian with a product state.

Another route to a state with a large overlap with the ground state, is to prepare the VSCF state, and then adiabatically evolve under a Hamiltonian that changes slowly from the VSCF Hamiltonian, to the full vibrational Hamiltonian $\Hamilt$. This approach has received significant attention within quantum computing approaches to the electronic structure problem, since it was first proposed in the context of quantum computational chemistry in Ref.~\cite{aspuru2005simulated}. However, both adiabatic state preparation and phase estimation typically require long circuits, with a large number of gates. As a result, quantum error correction is required to suppress the effect of device imperfections. It is therefore helpful to introduce variational methods, which may make these calculations feasible for near-term, non-error corrected quantum computers. Variational methods replace the long gate sequences required by phase estimation with a polynomial number of shorter circuits~\cite{peruzzo2014variational, VQETheoryNJP}. This dramatically reduces the coherence time required. As a result, quantum error correction may not be required, if the error rate is sufficiently low, in the context of the number of gates required. The circuits used consist of a number of parametrised gates which seek to create an accurate approximation of the desired state. The parameters are updated using a classical feedback loop, in order to produce better approximations of the desired state. The circuit used is known as the `ansatz' circuit.

Inspired by classical methods for the vibrational structure problem, we introduce the unitary vibrational coupled cluster (UVCC) ansatz. This is a unitary analogue of the VCC ansatz introduced in Refs.~\cite{christiansen2004vcc, christiansen2004secondquantised}. We note that a similar pairing exists for the electronic structure problem, where the unitary coupled cluster (UCC) ansatz~\cite{yung2014transistor,UCC} has been suggested as a quantum version of the classical coupled cluster method. The UVCC ansatz is given by
\begin{equation}
\ket{\Psi(\vec{\theta})} = \exp(\hat{T}-\hat{T}^\dag) \ket{\Psi_0},
\end{equation}
where the initial state $\ket{\Psi_0}$ can be either the ground state $\ket{\psi_0}$ of the harmonic oscillators or the VSCF state $\ket{\Psi_{\textrm{VSCF}}}$, $\hat{T}$ is the sum of molecular excitation operators truncated at a specified excitation rank, and $\vec{\theta}$ are the parameters defined below. Similar to the unitary coupled cluster in electronic structure problems, the single and double excitation operators are
\begin{equation}
	\hat{T} = \hat{T}_1 + \hat{T}_2 + \ldots,
\end{equation}
with
\begin{equation}
	\begin{aligned}
		\hat{T}_1 &= \sum_{m=1}^{M}\sum_{s_m,t_m=0}^{d-1}\theta_{s_m,t_m}\ketbra{s_m}{t_m},\\
		\hat{T}_2 &= \sum_{m<n}^{M}\sum_{s_m, t_m,p_n, q_n=0}^{d-1}\theta_{s_m,t_m,p_m,q_n}\ketbra{s_mp_n}{t_mq_n}.
	\end{aligned}
\end{equation}
Here, we omit the subscript of modes for simplicity. $\theta_{s_m,t_m}$ and $\theta_{s_m,t_m,p_m,q_n}$ are real parameters, and $\vec{\theta}=\{\theta_{s_m,t_m},\theta_{s_m,t_m,p_m,q_n}\}$. The $\hat{T}$ operators can be mapped to qubit operators via either the direct or compact mapping. 

The UVCC ansatz seeks to create a good approximation to the true ground state by considering excitations above a reference state. We note that the classical VCC ansatz is not a unitary operator. Correspondingly, the method is not variational, meaning that energies are not bounded from below. Moreover, we expect that the UVCC ansatz will deal better with problems of strong static correlation than the VCC ansatz, as the former can be used with multi-reference states. This echoes the way in which the UCC ansatz can be used with multi-reference states~\cite{UCC}, while it is typically more difficult when using the the canonical CC method~\cite{helgaker2014molecular}.)

Once we have obtained the energy levels of the vibrational Hamiltonian using the methods discussed above, we can calculate the infrared and Raman frequencies, using the difference between the excited and ground-state energies~\cite{wilson1980molecular}. 

It is often also the case that one is interested in the properties of a system in thermal equilibrium, rather than a specific eigenstate. We can also use established quantum algorithms with the Hamiltonians described above to construct these thermal states. On error corrected quantum computers, we can use the heuristic algorithms presented in Refs.~\cite{temme2011metropolis, yung2012metropolis} to construct these thermal states. Alternatively, we can use near-term devices to implement hybrid algorithms for imaginary time evolution~\cite{mcardle2018variational,yuan2018variational,motta2019imaginary}.

\subsection{Franck-Condon factors}\label{FCFSec}
In addition to focusing on the eigenstates or thermal states of a single vibrational Hamiltonian, we can also consider vibronic (vibrational and electronic) transitions between the vibrational levels resulting from different electronic potential energy surfaces. Consider two electronic states, $\ket{i}_e$ and $\ket{f}_e$. The molecular Hamiltonian is 
\begin{equation}
	H_{mol} = \ketbra{i}_e\otimes H_i+\ketbra{f}_e\otimes H_f,
\end{equation}
where $H_i$ and $H_f$ are vibrational Hamiltonians, with energy eigenstates $\ket{\psi^i_{vib}}$ and $\ket{\psi^f_{vib}}$, respectively. Using Fermi's Golden Rule, the probability of a photon-induced transition between two wavefunctions $\ket{\psi^i}=\ket{i}\otimes \ket{\psi^i_{vib}}$ and $\ket{\psi^f} = \ket{f}\otimes\ket{\psi^f_{vib}}$ is proportional to the square of the transition dipole moment $P^2 = \lvert\mel{\psi^f}{\hat{\mu}}{\psi^i}\rvert^2$, using first order time-dependent perturbation theory. Within the Condon approximation, $\hat{\mu}=\hat{\mu}_e + \hat{\mu}_N$, the transition probability becomes proportional to $P^2 = \left|\braket{\psi^f_{vib}}{\psi^i_{vib}}\right|^2\cdot\left|\mel{f}{\hat{\mu}_e}{i}\right|^2$. Here $\left|\braket{\psi^f_{vib}}{\psi^i_{vib}}\right|^2$ are referred to as Franck-Condon integrals. 
Without the Condon approximation, the Franck-Condon integrals become $\left|\bra{\psi^f_{vib}}\hat{\mu}(\Q)\ket{\psi^i_{vib}}\right|^2$, with $\hat{\mu}(\Q) = \left|\bra{f}\hat{\mu}\ket{i}\right|^2$. 

In practice, $\ket{\psi^i_{vib}}$ and $\ket{\psi^f_{vib}}$ are eigenstates of Hamiltonians with different harmonic oscillator normal modes $\Q^f$ and $\Q^i$. These modes are related by the Duschinsky transform $\Q^f = \mathbf{U}\Q^i+\darkd$~\cite{huhthesis,Kupka}. According to the Doktorov unitary representation of the Duschinsky transform, the harmonic oscillator eigenstates are related by~\cite{doktorov1977transitions,huh2015boson,huh2017vibronic}
\begin{equation}
\ket{\mathbf s_f} = \hat{U}_{Dok} \ket{\mathbf s_i}
\end{equation}
where $\ket{\mathbf s_i}$ and $\ket{\mathbf s_f}$ are harmonic oscillator eigenstates in the initial and final coordinates $\Q^i$ and $\Q^f$, respectively. The Doktorov unitary can be decomposed into a product of unitary operators $\U_{Dok} = \U_t \U_{s'}^\dag \U_s \U_r$, which depend on the displacement vector $\darkd$, the rotation matrix $\mathbf{U}$, and matrices of the frequencies of the harmonic oscillators $\mathbf{\Omega}^i = 1/\hbar~\mathrm{diag}(\sqrt{\omega^i})$ and $\mathbf{\Omega}^f = 1/\hbar~\mathrm{diag}(\sqrt{\omega^f})$. The definitions of the unitary operators are shown in Appendix~\ref{AppendixDusch}.

If $\ket{\Psi^i_{vib}}$ and $\ket{\Psi^f_{vib}}$ are the qubit wavefunctions resulting from diagonalisation of $H_i$ and $H_f$ using a quantum computer, they will be obtained in different normal mode bases $\ket{\mathbf s_i}$ and $\ket{\mathbf s_f}$, respectively. We cannot directly calculate the Franck-Condon integrals using $\left|\braket{\Psi^f_{vib}}{\Psi^i_{vib}}\right|^2$, as this does not take into account the different bases. Instead, we must implement the Doktorov unitary to get the Franck-Condon integrals 
\begin{equation}
	\left|\braket{\psi^f_{vib}}{\psi^i_{vib}}\right|^2 = \left|\bra{\Psi^f_{vib}}\U_{Dok}\ket{\Psi^i_{vib}}\right|^2.
\end{equation}
The Franck-Condon integrals without the Condon approximation can be efficiently calculated via
\begin{equation}
	\left|\bra{\psi^f_{vib}}\hat{\mu}(\Q)\ket{\psi^i_{vib}}\right|^2 = \left|\bra{\Psi^f_{vib}}\hat{\mu}(\Q^f)\U_{Dok}\ket{\Psi^i_{vib}}\right|^2.
\end{equation}
They can be both efficiently calculated with the generalised {\small SWAP}-test circuit~\cite{SWAPOverlap}.

Alternatively, we can obtain the Franck-Condon integrals without realising the Doktorov transform. The qubit states $\ket{\Psi^i_{vib}}$ and $\ket{\Psi^f_{vib}}$ are obtained from $H_i(\Q^i)$ and $H_f(\Q^f)$ with normal mode coordinates $\Q^i$ and $\Q^f$, respectively. Instead, we can focus on one set of normal mode coordinates $\Q^i$ and represent the Hamiltonian $H_f$ in $\Q^i$, $H_f'(\Q^i)$. By solving the energy eigenstates of $H_f'(\Q^i)$, we can directly get $\ket{\Psi^{'f}_{vib}} = \U_{Dok}^\dag\ket{\Psi^f_{vib}}$ and calculate the Franck-Condon integrals without realising the Doktorov transform. However, as the Hamiltonian $H_f'(\Q^i)$ is not encoded in the correct normal mode basis, the ground state of the harmonic oscillators or the VSCF state $\ket{\Psi_{\textrm{VSCF}}}$ may not be an ideal initial state to start with. However, this effect may be negligible if the overlap between $\ket{\Psi^i_{vib}}$ and $\ket{\Psi^{'f}_{vib}}$ is suitably large. In this case, the initial state $\ket{\Psi_0}$ for $\ket{\Psi^i_{vib}}$ should also be an ideal initial state for $\ket{\Psi^{'f}_{vib}}$. The aforementioned transformation can be implemented by transforming the normal mode coordinates $\Q^i$ as described in Ref.~\cite{huhthesis}.


\subsection{Vibrational dynamics}
In this section, we consider methods to investigate the dynamic properties of vibrational Hamiltonians. Vibrational dynamics underpin phenomena including energy and electron transport~\cite{ElectronTransferRates1,ElectronTransferRates2} and chemical reactions~\cite{VibrExcitation, ProteinVibr, Proctor2008}. Dynamical behaviour can be studied by transforming to a single-mode basis of spatially localised vibrational modes, as described in Ref.~\cite{sparrow2018simulating}. The spatially localised vibrational modes $a^{L}_i$, are related to the normal modes $a_i$ via a basis transformation
\begin{equation}
	a_i = \sum_{i,j}U_{i,j}a_{j}^L,
\end{equation}
with real unitary matrix $U_{i,j}$. We can obtain the corresponding localised Hamiltonian $H^L$, using the transformation of the normal coordinates and momenta
\begin{equation}\label{Eq:transform}
	  p_i=\sum_{\red{j}}\sqrt{\frac{\omega_i}{\omega_j}}U_{i,j}p_{j}^L,\quad
	  q_i=\sum_{\red{j}}\sqrt{\frac{\omega_j}{\omega_i}}U_{i,j}q_{j}^L.
\end{equation}
Given an initial state of the localised vibrations, the dynamics can be simulated by applying the time evolution operator $e^{-iH^L t}$. This can be achieved in a number of ways, using different Hamiltonian simulation algorithms, including: Trotterization (also referred to as product formulae)~\cite{trotter1959product,lloyd1996universal}, the Taylor series method~\cite{berry2012black,PhysRevLett.114.090502,Berry15optimal}, and qubitization~\cite{low2016hamiltonian, low2018hamiltonian} in conjunction with quantum signal processing~\cite{LowQSPprx, PhysRevLett.118.010501}. The product formula method is the most simple to realise. If $H^L$ can be decomposed as $H^L =  \sum_i \lambda_i^L {h}_i^L$, the time evolution operator $e^{-iH^L t}$ can be realised using a product formula,
\begin{equation}
    e^{-iH^L t} = \left(\prod_i e^{-i\lambda_i^LH_i^L t/N}\right)^N + O(t^2/N)
\end{equation}
when $N$ is chosen to be sufficiently large to suppress the error in the approximation.

Alternatively, the vibrational dynamics can be realised using a recently proposed variational algorithm~\cite{Li2017}. One could use either a UVCC ansatz, or a Trotterized ansatz~\cite{PhysRevA.95.032338, jones2018compilation}.

\section{Numerical simulations}\label{Sec:simulation}
In this section, we demonstrate how the techniques described above can be used to calculate the vibrational energy levels of small molecules. We focus on the polyatomic molecules \ch{H2O} and \ch{SO2}, which both have three vibrational modes. We considered both nuclear potentials expanded to fourth order, the coefficients of which are shown in Table \ref{tab:coe} in Appendix~\ref{AppendixNumerics}. We consider the cases with two and four energy levels for each of the harmonic oscillator modes, yielding Hamiltonians acting on 6 and 12 qubits for the direct mapping, and 3 and 6 qubits for the compact mapping. We used the compact mapping in our numerical simulations, as it requires fewer qubits. There are 216 and 165 terms in the Hamiltonian for \ch{H2O} and \ch{SO2}, respectively. 

\begin{figure}[t]\centering
  \includegraphics[width=\linewidth]{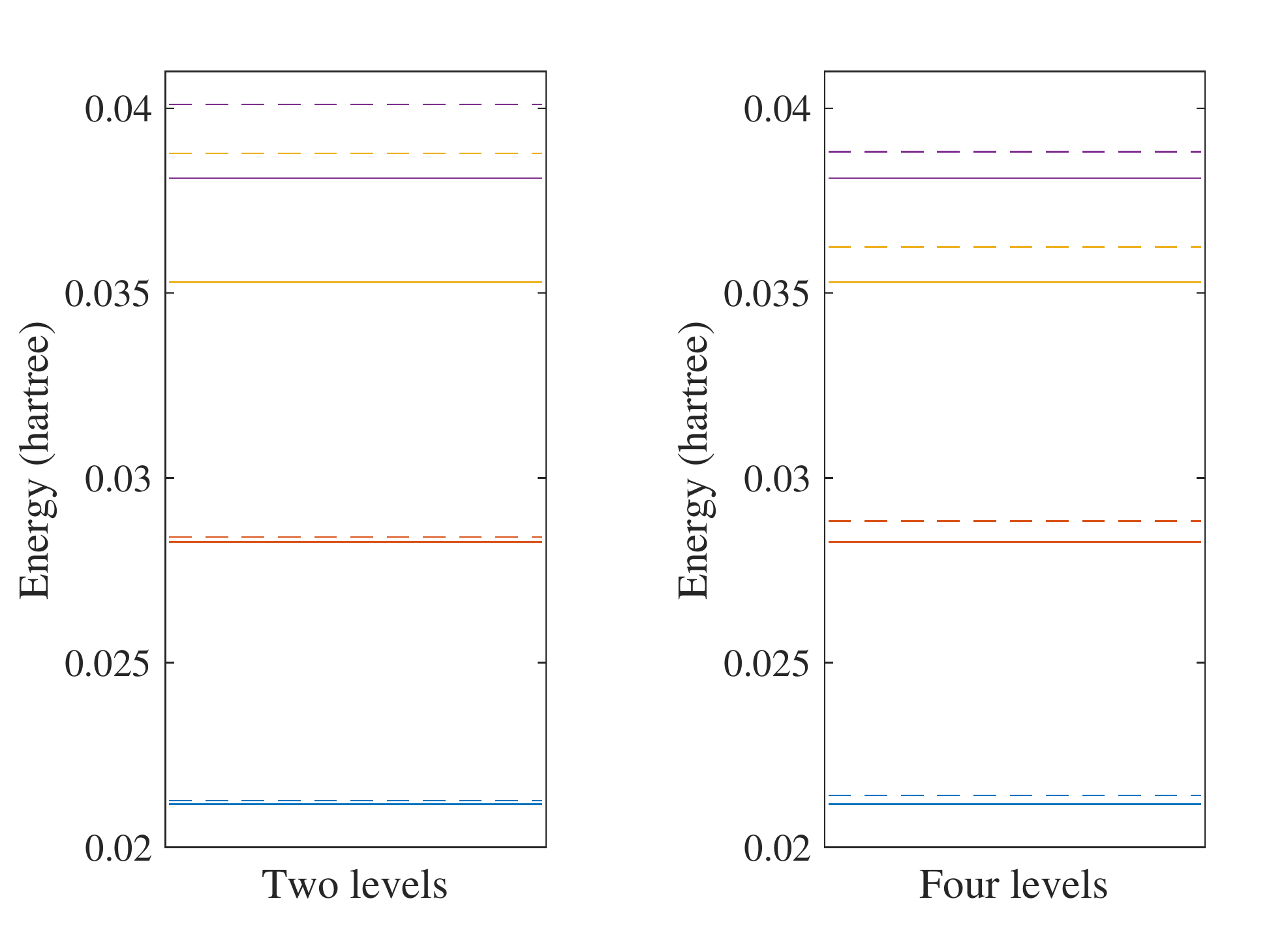}
  \caption{Vibrational spectra of \ch{H2O} with two and four energy levels for each mode. The solid lines are the energy levels of the harmonic oscillator eigenstates and the dashed lines are the vibrational spectra  of the Hamiltonian with a fourth order expansion of the potential. }\label{Fig:res}
\end{figure}

We first calculate the energy levels under the harmonic approximation. We compare this to the energy levels obtained with a fourth order expansion of the potential. The results for \ch{H2O} are shown in Fig.~\ref{Fig:res}. We can see that although the ground state can be well approximated by the harmonic oscillators, the excited states deviate from the harmonic oscillators at higher energy levels. The results for \ch{SO2} can be found in the Appendix. These calculations highlight the importance of anharmonic terms in the potential for even small molecules.

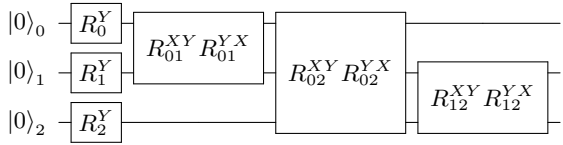
\begin{figure}[t]
\centering
\begin{align*}
\Qcircuit @C=.5em @R=.4em {
\lstick{\ket{0}_0}&\gate{R^Y_0}&\multigate{1}{R^{XY}_{01}R^{YX}_{01}} &\multigate{2}{R^{XY}_{02}R^{YX}_{02}} &\qw&\qw\\
\lstick{\ket{0}_1}&\gate{R^Y_1}&\ghost{R^{XY}R^{XY}}&\ghost{R^{XY}_{02}R^{YX}_{02}}&\multigate{1}{R^{XY}_{12}R^{YX}_{12}}&\qw\\
\lstick{\ket{0}_2}&\gate{R^Y_2}&\qw&\ghost{R^{XY}R^{YX}} &\ghost{R^{XY}R^{YX}}&\qw\\
}
\end{align*}
\caption{The UVCC ansatz of three modes each with two energy levels. There are nine gates with six parameters (joined gates share the same parameters). \red{The single qubit gate on the $i^{\textrm{th}}$ qubit is $R^Y_i(\theta) = e^{-i\theta\sigma_y^i}$, and the two qubit gate on the $i^{\textrm{th}}$ and $j^{\textrm{th}}$ is $R^{AB}_{ij}(\theta) = e^{-i\theta(\sigma_A^i\otimes\sigma_B^j)}$.}} \label{Fig:circuit}
\end{figure}

Next, we implemented the UVCC ansatz to obtain the vibrational energy levels of \ch{H2O} using the variational quantum eigensolver~\cite{peruzzo2014variational}. For simplicity, we considered two energy levels for each mode. 
To implement the UVCC ansatz, we first calculate the imaginary part of $\hat T$ and encode it into a linear combination of local Pauli terms, i.e., 
\begin{equation}
    \hat T - \hat T^\dag = i\sum_i \alpha_i(\theta) \sigma_i. 
\end{equation}
Then, as for the UCC ansatz, we realise $\exp(\hat T - \hat T^\dag)$ by a first order Trotterisation via $\prod_i \exp(i \alpha_i(\theta) \sigma_i)$. For example, the UVCC ansatz of \ch{H2O} with two energy levels can be prepared by the circuit in Fig.~\ref{Fig:circuit}.  

\begin{figure}[t]\centering
  \includegraphics[width=\linewidth]{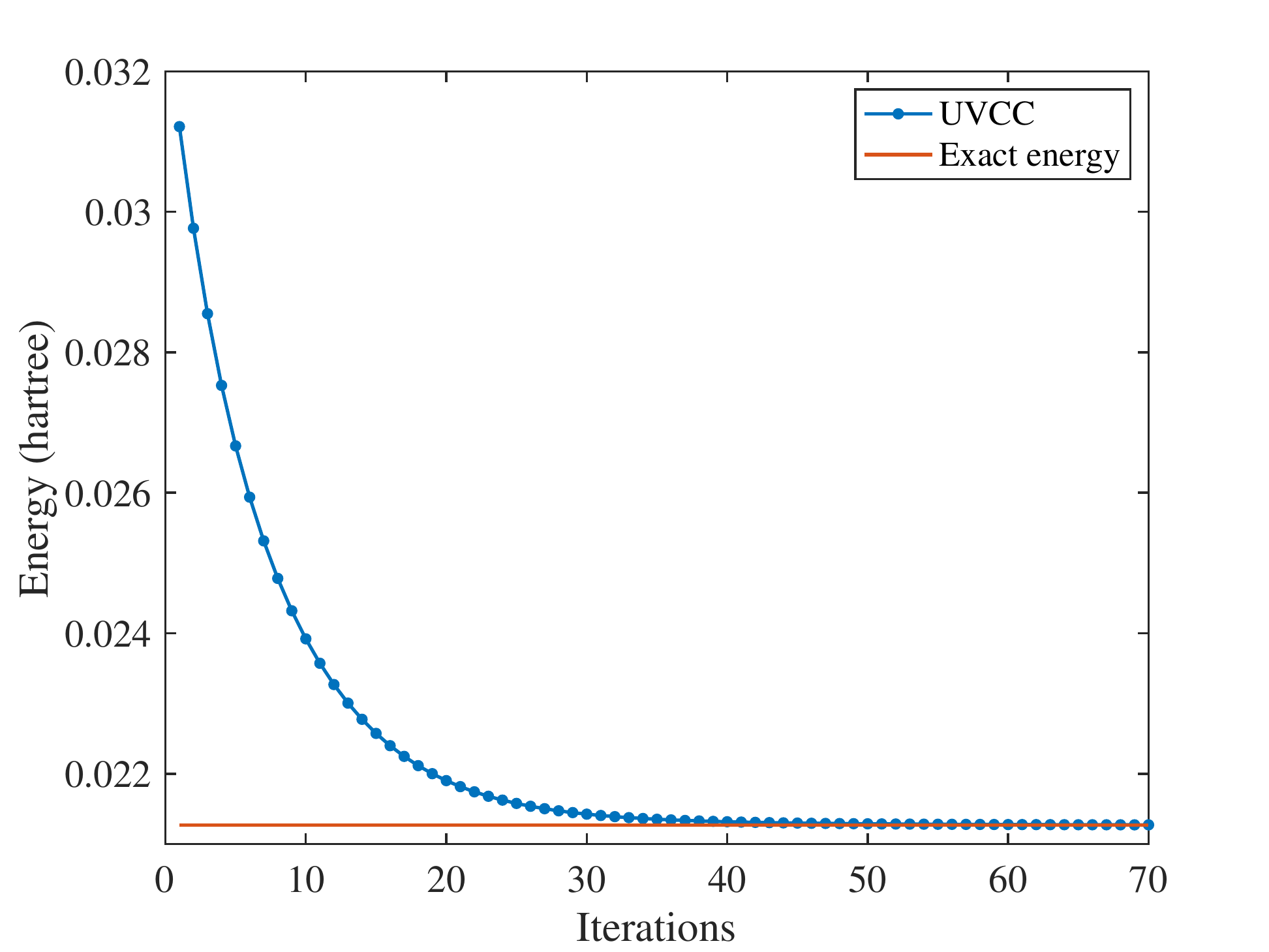}
  \caption{Solving the vibrational ground state of \ch{H2O} with the UVCC ansatz. Here, we consider two energy levels for each mode.  }\label{Fig:resh2o}
\end{figure}

Using the UVCC ansatz, we can obtain the vibrational ground state with a variational procedure. As the ground state is close to the initial state $\ket{0}^{\otimes 3}$, we start with parameters slightly perturbed from zero. We then use gradient descent to find the minimum energy of the system. The results are shown in Fig.~\ref{Fig:resh2o}.

\section{Discussion}\label{Sec:Discussion}
In this work, we have extended many of the techniques developed for quantum simulation to the problem of simulating vibrations. Understanding vibrations is an important problem for accurately modelling chemical systems, yet one that is difficult to solve classically. 

We have discussed ways to map between vibrational modes and qubit states. This is only possible when we consider a restricted number of harmonic oscillator energy levels. This approximation is appropriate when investigating the low energy properties of the Hamiltonian, such as the ground state energy. However, it may not always be possible to use this approximation when considering time evolution, as this requires the exponentiation of our truncated Hamiltonian. One possible route to overcome this challenge is to repeat simulations which consider an increasing number of energy levels, and then to extrapolate to the infinite energy level result. This technique was used in a similar context to this work in Ref.~\cite{dumitrescu2018cloud}.

Once the vibrational Hamiltonian has been mapped to a qubit Hamiltonian, much of the existing machinery for quantum simulation can be applied. Static properties, such as energy levels, can be calculated using phase estimation or variational approaches. To aid variational state preparation, we proposed a unitary version of the powerful VCC method used in classical vibrational simulations. The resulting energy eigenstates can be used as an input for {\small SWAP}-test circuits. These calculate the Franck-Condon factors for the molecules, which are related to the absorption spectra. Alternatively, one may investigate dynamic properties, using methods for Hamiltonian simulation to time evolve a specified state.

Compared with analog algorithms \cite{huh2015boson,huh2017vibronic,shen2018vibronic, sparrow2018simulating}, our method can easily take into account anharmonic terms in the nuclear Hamiltonian. Moreover, it could be used to simulate large systems by protecting the quantum computer with error correction. As our technique is tailored for simulating vibrational states, it makes it simple to investigate interesting vibrational properties, such as the Franck-Condon factors. ~\red{However, as our approach uses the Born-Oppenheimer approximation, it is not suitable for all problems in chemistry, such as problems including relativistic effects~\cite{reiher2014relativistic} or conical intersections~\cite{domcke2004conical, domcke2012spectroscopy, ryabinkin2017conical}.} Future work will address whether restricting our vibrational modes to low-lying energy levels poses a significant challenge for problems of practical interest. \\

\section*{Note added}
After this work was released as a preprint, a notable related work was released online~\cite{sawaya2018vibronic}. That paper provides a quantum algorithm for efficiently calculating the entire converged Franck-Condon profile, as opposed to the method we present herein, which calculates a single Franck-Condon factor. However, the techniques we have discussed for finding vibrational ground states and thermal states provide a way to realise the crucial first step of their algorithm. As such, our two works are highly complementary, and can be compared. \\

\section*{Acknowledgements.}
We acknowledge insightful comments from Joonsuk Huh.
This work was supported by BP plc and the EPSRC National Quantum Technology Hub in Networked Quantum Information Technology (EP/M013243/1). X. Shan acknowledges the use of the University of Oxford Advanced Research Computing (ARC) facility.

\bibliographystyle{apsrev4-1}
\bibliography{ChemistryReviewBib}

\clearpage
\widetext
\newpage
\appendix
\section{Encoding vibrational Hamiltonians into qubits}\label{ProblemMapping}
The molecular Hamiltonian in atomic units is
\begin{equation}
\begin{aligned}
	\Hamilt_{mol} =& -\sum_i\frac{\nabla^2_i}{2}  -\sum_I\frac{\nabla^2_I}{2M_I} - \sum_{i,I}\frac{Z_I}{|\mathbf{r}_i-\mathbf{R}_I|}\\
	&+\frac{1}{2}\sum_{i\neq j}\frac{1}{|\mathbf{r}_i-\mathbf{r}_j|}+\frac{1}{2}\sum_{I\neq J}\frac{Z_IZ_J}{|\mathbf{R}_I-\mathbf{R}_J|}.
\end{aligned}
\end{equation}where $M_I$, $\mathbf{R}_I$, and $Z_I$ are the mass, position, and charge of nuclei $I$, and $\mathbf{r}_i$ is the position of electron $i$. Given the location of the nucleus, the electronic Hamiltonian is
\begin{equation}
\begin{aligned}
	\Hamilt_e(\mathbf{R}_I) &=  -\sum_i\frac{\nabla^2_i}{2} + \sum_{i,I}\frac{Z_I}{|\mathbf{r}_i-\mathbf{R}_I|}+\frac{1}{2}\sum_{i\neq j}\frac{1}{|\mathbf{r}_i-\mathbf{r}_j|},
\end{aligned}
\end{equation}
and the total Hamiltonian is
\begin{equation}
\Hamilt_{mol} =  -\sum_I\frac{\nabla^2_I}{2M_I} +\frac{1}{2}\sum_{I\neq J}\frac{Z_IZ_J}{|\mathbf{R}_I-\mathbf{R}_J|} + \Hamilt_e(\mathbf{R}_I).
\end{equation}
Under the Born-Oppenheimer approximation, we assume the electrons and nuclei are in a product state, 
\begin{equation}
	\ket{\psi} = \ket{\psi}_n\ket{\psi}_e.
\end{equation}
To get the ground state of the Hamiltonian, one can thus separately minimise over $\ket{\psi}_n$ and $\ket{\psi}_e$,
\begin{equation}
E_0 = \min_{\ket{\psi}_n}\min_{\ket{\psi}_e} \bra{\psi}_n\bra{\psi}_e \Hamilt_{mol} \ket{\psi}_n\ket{\psi}_e.
\end{equation}
As only $\Hamilt_e(\mathbf{R}_I)$ depends on $\ket{\psi}_e$, the minimisation over  $\ket{\psi}_e$ is equivalent to finding the ground state of $\Hamilt_e(\mathbf{R}_I)$. Denote 
\begin{equation}
V_{0}^e(\mathbf{R}_I) = \min_{\ket{\psi}_e} \bra{\psi}_e \Hamilt_e(\mathbf{R}_I) \ket{\psi}_e, 
\end{equation}
then the ground state of $\Hamilt_{mol}$ can be found by solving the ground state of $H_0$,
\begin{equation}
\Hamilt_0 = -\sum_I\frac{\nabla^2_I}{2M_I} +\frac{1}{2}\sum_{I\neq J}\frac{Z_IZ_J}{|\mathbf{R}_I-\mathbf{R}_J|} + V_{0}^e(\mathbf{R}_I).
\end{equation}

In general, considering a spectral decomposition of $\Hamilt_e(\mathbf{R}_I) = \sum_{s}V_{s}^e(\mathbf{R}_I)\ketbra{\psi_s}_e$, the molecular Hamiltonian is
\begin{equation}
	\Hamilt_{mol} = \sum_s\ketbra{\psi_s}_e\otimes H_s.
\end{equation}
Here, $\ket{\psi_s}_e$ are eigenstates of the electronic Hamiltonian and 
\begin{equation}
	\Hamilt_s = \sum_I - \frac{\nabla^2_I}{2M_I} +V_s(\mathbf{R}_I),
\end{equation}
and
\begin{equation}
	 V_s(\mathbf{R}_I) = \frac{1}{2}\sum_{I\neq J}\frac{Z_IZ_J}{|\mathbf{R}_I-\mathbf{R}_J|} + V_{s}^e(\mathbf{R}_I).
\end{equation}

Finding the spectra of $\Hamilt_e(\mathbf{R}_I) $ is called the electronic structure problem, which can be efficiently solved using a quantum computer~\cite{aspuru2005simulated}. One approach is to consider a subspace that the ground state lies in and transform the Hamiltonian $\Hamilt_e(\mathbf{R}_I) $ into the second quantised formulation, with a basis determined by the subspace. As electrons are fermions, the obtained Hamiltonian is a fermionic Hamiltonian. By using the standard encoding methods, such as Jordan-Wigner and Bravyi-Kitaev~\cite{BravyiKitaev12}, the fermionic Hamiltonian is converted into a qubit Hamiltonian, whose spectra can be efficiently computed. 

Focusing on the ground state of the electronic structure Hamiltonian, we show how to redefine $\Hamilt_0$ in the mass-weighted basis and how to encode it with qubits. Denote $V = \frac{1}{2}\sum_{I\neq J}\frac{Z_IZ_J}{|\mathbf{R}_I-\mathbf{R}_J|} + V^e_0(\mathbf{R}_I)$, then one can obtain the mass-weighted normal coordinates $\Qsingle_i$ by minimising the coupling between the rotational and vibrational degrees of freedom and diagonalising the Hessian matrix, 
\begin{equation}
H_{IJ} = \frac{1}{\sqrt{M_IM_J}}\frac{\partial^2V}{\partial \mathbf{R}_I\partial \mathbf{R}_J}.
\end{equation}
In the mass-weighted basis, the potential can be expanded via a Taylor series truncated at fourth order
\begin{equation}
V^{(4)} = \frac{1}{2}\sum_i \omega_i^2 \Qsingle_i^2 + \sum_{i\leq j\leq k} f_{ijk}\Qsingle_i\Qsingle_j\Qsingle_k + \sum_{i\leq j\leq k\leq l} f_{ijkl} \Qsingle_i\Qsingle_j\Qsingle_k\Qsingle_l,
\end{equation}
and the total Hamiltonian becomes
\begin{equation}
\Hamilt_0 =  \sum_i - \frac{\nabla^2_i}{2} +V^{(4)}.
\end{equation}
If we consider the higher order terms as a perturbation, one can get the normal modes by solving the harmonic oscillator 
\begin{equation}
\hat{h}_i =  -\frac{\nabla^2_i}{2}  +\frac{1}{2} \omega_i^2 \Qsingle_i^2.
\end{equation}
We denote the eigenbasis for $\hat h_i$ as $\{\ket{\psi^i_{s^i}},\forall s^i = 0, 1, \dots\}$, then  the nuclear wave function can be represented by
\begin{equation}
\ket{\psi} = \sum_{s^1s^2\dots s^M}\alpha_{s^1s^2\dots s^M}\ket{\psi^1_{s^1}\psi^2_{s^2}\dots\psi^M_{s^M}}.
\end{equation}
The Hamiltonian under the normal mode basis becomes,
\begin{equation}
H_{s^1s^2\dots s^M,t^1t^2\dots t^M} = \bra{\psi^1_{s^1}\psi^2_{s^2}\dots\psi^M_{s^M}}\Hamilt_0\ket{\psi^1_{t^1}\psi^2_{t^2}\dots\psi^M_{t^M}}.
\end{equation}Suppose the basis $\{\ket{\psi^i_{s^i}},\forall s^i = 0, 1, \dots\}$ is truncated to the lowest $d$ energy levels, the space of $H_{s^1s^2\dots s^M,t^1t^2\dots t^M}$ is equivalent to the space of $M$ $d$-level systems, or equivalently $M\log_2d$ qubits.


\section{Variational quantum simulation with the unitary vibrational coupled cluster ansatz}

We can make use of variational methods to find the low energy spectra of the vibrational Hamiltonian. Inspired by classical computational chemistry, we introduce the unitary vibrational coupled cluster (UVCC) ansatz
\begin{equation}
\ket{\text{VCC}} = \exp(\hat{T}-\hat{T}^\dag) \ket{\Phi_0}
\end{equation}
where the reference state $\ket{\Phi_i}$ is a properly chosen initial state, and $T$ is the sum of molecular excitation operators truncated at a specified excitation rank (often single and double excitations)~\cite{UCC}
\[\hat{T} = \hat{T}_1 + \hat{T}_2 + \ldots.\]
with

\[\hat{T}_1 = \sum_{m}^{M}\sum_{a^m,i^m}t_{a^m,i^m}^{m}\ketbra{\psi^m_{a^m}}{\psi^m_{i^m}},\]
\[\hat{T}_2 = \sum_{m<n}^{M}\sum_{a^m, i^m}\sum_{a^n, j^n}t_{a^m,a^n,i^m,j^n}^{m,n}\ketbra{\psi^m_{a^m}\psi^n_{a^n}}{\psi^m_{i^m}\psi^n_{i^n}}.\]

The initial state can be the product of the ground-state of each mode
\begin{equation}
\ket{\Phi_0} = \ket{\psi^1_{0}\psi^2_{0}\dots\psi^M_{0}}.
\end{equation}
Or we can also run a vibrational self-consistent field (VSCF) to get the Hartree-Fock initial state
\begin{equation}
\ket{\Phi_0} = \ket{\phi^1\phi^2\dots\phi^M},
\end{equation}
which is obtained by minimising the energy of the Hamiltonian
\begin{equation}
\min_{\ket{\phi^1\phi^2\dots\phi^M}} \bra{\phi^1\phi^2\dots\phi^M}\Hamilt_0\ket{\phi^1\phi^2\dots\phi^M}
\end{equation}
by solving the self-consistent equation
\begin{equation}
 \Hamilt_i\ket{\phi_i} = E_i\ket{\phi_i},
\end{equation}
with $\Hamilt_i = \bra{\phi^1\dots\phi^{i-1}\phi^{i+1}\dots\phi^M}\Hamilt_0\ket{\phi^1\dots\phi^{i-1}\phi^{i+1}\dots\phi^M}$.

\section{Duschinsky transform}\label{AppendixDusch}
The relation between the initial coordinates $\Qsingle_1$ and final coordinates $\Qsingle_2$ is
\begin{equation}
\begin{aligned}
	\Q_1 &= \mathbf{U}\Q_2 + \mathbf{d},\\
\end{aligned}
\end{equation}
where $\mathbf{U}$ is the Duschinsky rotation matrix and $\mathbf{d}$ is the displacement vector. 
The harmonic oscillator Hamiltonian with unit mass is
\begin{equation}
	\hat{h}_i = -\frac{{p}_i^2}{2}  +\frac{1}{2} \omega_i^2 \Qsingle_i^2,
\end{equation}
with ${{p}}_i = \frac{\partial }{\partial {q_i}}$. The operators ${\mathbf{\Qsingle}}_1 $ and $\mathbf{{p}}_1$ can be represented by the creation and annihilation operators 
\begin{equation}
	\begin{aligned}
		\Q_1 = \sqrt{\frac{\hbar}{2\mathbf{\Omega_1}}}\left(\mathbf{\annih}_1 + \mathbf{a}^\dag_1\right),\,
		{\mathbf{p}}_1 = i\sqrt{\frac{\hbar\mathbf{\Omega_1}}{2}}\left(\textbf{a}_1-\mathbf{a}^\dag_1\right).
	\end{aligned}
\end{equation}
The transformation for ${\mathbf{p}}$ is ${\mathbf{p}}_1 = \mathbf{U}^\dag{\mathbf{p}}_2$. The transformation for the creation operators are
\begin{equation}
\begin{aligned}
	\mathbf{a}_1^\dag &= \sqrt{\frac{\mathbf{\Omega}_1}{2\hbar}}\left(\mathbf{q}_1-\frac{i}{\mathbf{\Omega}_1}\mathbf{p}_1\right),\\
	&= \sqrt{\frac{\mathbf{\Omega}_1}{2\hbar}}\left(U\mathbf{q}_2 + \mathbf{d}-\frac{i}{\mathbf{\Omega}_1}U^\dag\mathbf{p}_2\right),\\
	&= \sqrt{\frac{\mathbf{\Omega}_1}{2\hbar}}\left(U\sqrt{\frac{\hbar}{2\mathbf{\Omega_2}}}\left(\mathbf{a}_2 + \mathbf{a}_2^\dag\right) -\frac{i}{\mathbf{\Omega}_1}U^\dag i\sqrt{\frac{\hbar\mathbf{\Omega_2}}{2}}\left(\mathbf{a}_2^\dag - \mathbf{a}_2\right)\right) + \sqrt{\frac{\mathbf{\Omega}_1}{2\hbar}}\mathbf{d},\\
		&= \frac{1}{2}\sqrt{\mathbf{\Omega_1}}U\sqrt{\mathbf{\Omega_2}}^{-1}\left(\mathbf{a}_2 + \mathbf{a}_2^\dag\right) + \frac{1}{2}\sqrt{\mathbf{\Omega_1}}^{-1}U^\dag\sqrt{\mathbf{\Omega_2}}\left(\mathbf{a}_2^\dag - \mathbf{a}_2\right) + \sqrt{\frac{\mathbf{\Omega}_1}{2\hbar}}\mathbf{d},\\
&= \frac{1}{2}(J-J')\mathbf{a}_2 + \frac{1}{2}(J+J')\mathbf{a}_2^\dag  + \sqrt{\frac{\mathbf{\Omega}_1}{2\hbar}}\mathbf{d},\\
\end{aligned}
\end{equation}
where $J = \sqrt{\mathbf{\Omega_1}}U\sqrt{\mathbf{\Omega_2}}^{-1}$ and $J' = \sqrt{\mathbf{\Omega_1}}^{-1}U^\dag\sqrt{\mathbf{\Omega_2}}$. \\

It was shown by Doktorov \textit{et al.}~\cite{doktorov1977transitions} that the Duschinsky transform can be implemented using a unitary transform inserted into the overlap integral
\begin{equation}
\bra{\nu_f} \ket{\nu_i} = \bra{\nu'_f} \hat{U}_{Dok} \ket{\nu_i}
\end{equation}
where $\ket{\nu}$ is a harmonic oscillator eigenstate in the initial coordinate $\Q$ and $\ket{\nu'}$ is a harmonic oscillator eigenstate in the final coordinate $\Q'$. The Doktorov unitary can be decomposed into a product of unitaries, which depend on the displacement vector $\vec{d}$, the rotation matrix $\mathbf{U}$, and matrices of the eigenenergies of the harmonic oscillator states ($\epsilon_i$, $\epsilon_i'$); $\mathbf{\Omega} = 1/\hbar~\mathrm{diag}(\sqrt{\epsilon_i})$ and $\mathbf{\Omega'} = 1/\hbar~\mathrm{diag}(\sqrt{\epsilon_i'})$. It is given by
\begin{equation}
\hat{U}_{Dok} = \hat{U}_t \hat{U}_{s'}^\dag \hat{U}_s \hat{U}_r
\end{equation}
where
\begin{equation}
\hat{U}_{t} = \mathrm{exp}\left(\frac{1}{2 \hbar} \vec{d}^t \mathbf{\Omega '} (\vec{a}^\dag - \vec{a})\right)
\end{equation}
where $\vec{a} = (a_0 , ..., a_M)^T$, and
\begin{equation}
\hat{U}_{s'} = \mathrm{exp}\left(-\frac{1}{2} (\vec{a}^\dag + \vec{a})^t \mathrm{ln}(\mathbf{\Omega '}) (\vec{a}^\dag - \vec{a}) + \frac{1}{2} \mathrm{Tr}(\mathrm{ln}(\mathbf{\Omega '}))\right)
\end{equation}
and
\begin{equation}
\hat{U}_{s} = \mathrm{exp}\left(-\frac{1}{2} (\vec{a}^\dag + \vec{a})^t \mathrm{ln}(\mathbf{\Omega}) (\vec{a}^\dag - \vec{a}) + \frac{1}{2} \mathrm{Tr}(\mathrm{ln}(\mathbf{\Omega}))\right)
\end{equation}
and
\begin{equation}
\hat{U}_{r} = \mathrm{exp}\left(\frac{1}{2} ((\vec{a}^\dag)^t \mathrm{ln}(\mathbf{U}) \vec{a} - (\vec{a})^t \mathrm{ln}(\mathbf{U}) \vec{a}^\dag\right).
\end{equation}

These exponentials could be expanded into local qubit operators using Trotterization. It is important to note that these relations are only valid when the single-mode basis functions are chosen to be harmonic oscillator eigenstates.



\section{Numerical simulation}\label{AppendixNumerics}

\begin{figure}[t]\centering
  \includegraphics[width=0.5\linewidth]{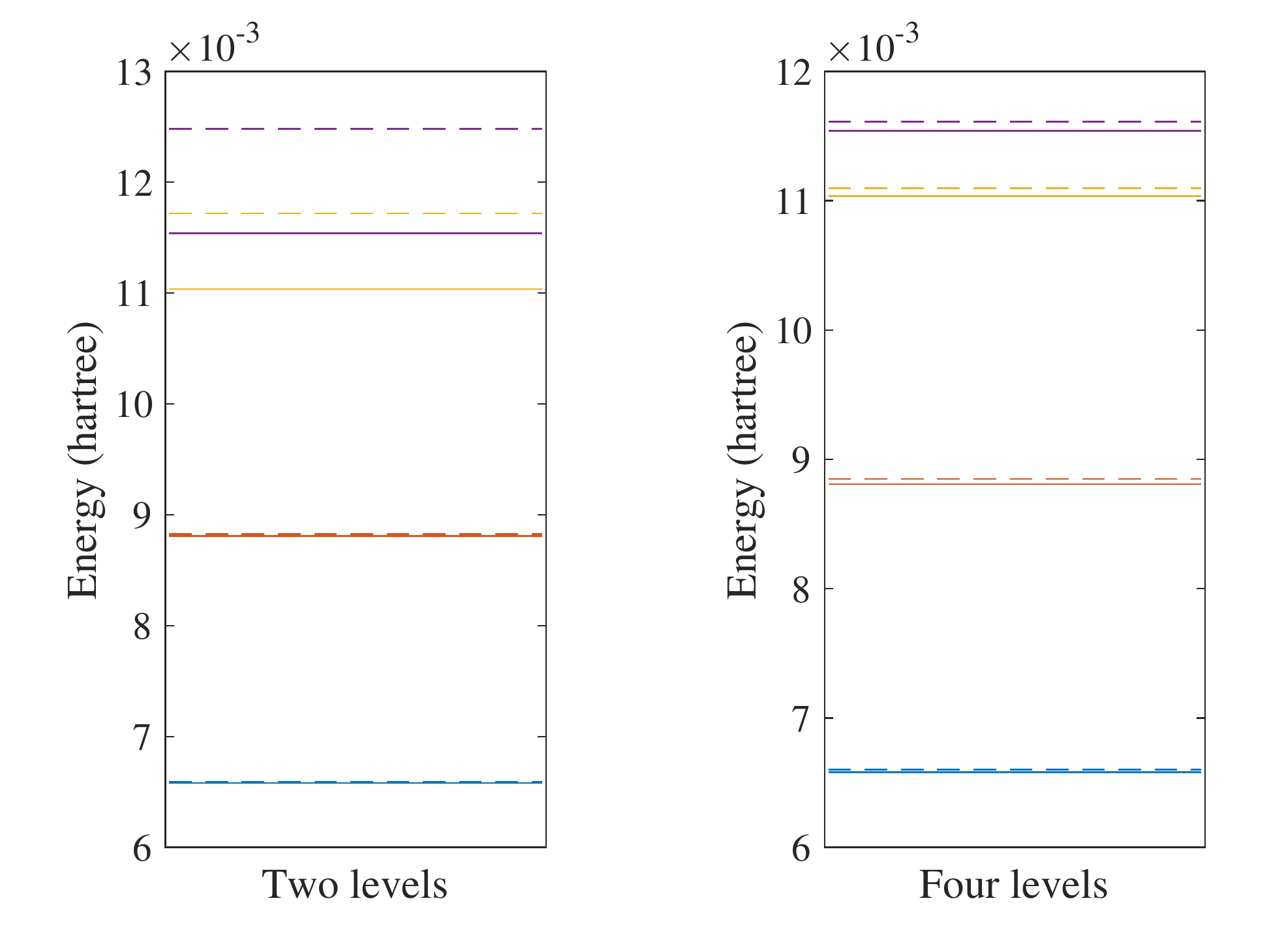}
  \caption{Vibrational spectra of \ch{SO2} with two and four energy levels for each mode. The solid lines are the energy levels of the harmonic oscillator eigenstates and the dashed lines are the vibrational spectra  of the Hamiltonian with a fourth order expansion of the potential. }\label{Fig:resSO2}
\end{figure}

The coefficients of \ch{H2O} and \ch{SO2} are shown in Table~\ref{tab:coe}, which were computed at MP2/aug-cc-pVTZ level of theory using Gaussian09 software \cite{frisch2009gaussian09}. The simulation results of the vibrational energy levels of \ch{SO2} are shown in Fig.~\ref{Fig:resSO2}.

\begin{table}[ht]
\centering
  \caption{Coefficients of the potential energy surface of \ch{H2O} and \ch{SO2}. The coefficients are in the atomic units, where the unit of length is $a_0 = 1$~Bohr ($0.529167~\times~10^{-10}$~m), the unit of mass is the electron mass $m_e$, and the unit of energy is 1~Hartree ($1~\textrm{Hartree}~= ~{e^2}/{4\pi\epsilon_0a_0}~=~27.2113~\textrm{eV}$). }
\begin{tabular}{ccc}
\hline
$k$&\ch{H2O}&\ch{SO2}\\
\hline
  $k_{1, 1}$ &$0.275240\times10^{-4}$& $0.252559\times10^{-5} $\\
  $k_{2, 2}$ &$  0.151618\times10^{-3}$&$0.125410\times10^{-4}$\\
  $k_{3, 3}$ &$ 0.161766\times10^{-3}$&$0.176908\times10^{-4} $\\
  $k_{1, 1, 1}$ &$ 0.121631\times10^{-6}$&$0.316646\times10^{-8}$\\
  $k_{1, 1, 2}$ &$ 0.698476\times10^{-6}$&$0.575325\times10^{-8}$\\
  $k_{1, 2, 2}$ &$ -0.266427\times10^{-6} $&$0.197771\times10^{-7}$\\
  $k_{2, 2, 2}$ &$ -0.312538\times10^{-5} $&$-0.668689\times10^{-7}$\\
  $k_{1, 3, 3}$ &$ -0.915428\times10^{-6} $&$-0.370850\times10^{-9}$\\
  $k_{2, 3, 3}$ &$ -0.964649\times10^{-5} $&$-0.284244\times10^{-6}$\\
  $k_{1, 1, 1, 1}$ &$ -0.463748\times10^{-9} $&$0.330842\times10^{-11}$\\
  $k_{1, 1, 2, 2}$ &$ -0.449480\times10^{-7} $&$-0.172869\times10^{-9}$\\
  $k_{1, 2, 2, 2}$ &$ 0.957558\times10^{-8} $&$-0.215928\times10^{-9}$\\
  $k_{2, 2, 2, 2}$ &$ 0.433267\times10^{-7} $&$0.225400\times10^{-9} $\\
  $k_{1, 1, 3, 3}$ &$ -0.555026\times10^{-7} $&$-0.356155\times10^{-9} $\\
  $k_{1, 2, 3, 3}$ &$ 0.563566\times10^{-7} $&$ -0.128135\times10^{-9} $\\
  $k_{2, 2, 3, 3}$ &$ 0.269239\times10^{-6} $&$0.220168\times10^{-8}$\\
  $k_{3, 3, 3, 3}$ &$ 0.462143\times10^{-7} $&$0.458046\times10^{-9}$\\
  $k_{2, 3, 3, 3}$ &0 &$-0.720760\times10^{-11}$\\
\hline
\end{tabular}
\label{tab:coe}
\end{table}

\end{document}